# Unexpected thermo-elastic effects in liquid glycerol by mechanical deformation

Eni Kume[1], Alessio Zaccone[2,3], Laurence Noirez[1]

1 Laboratoire Léon Brillouin (CEA-CNRS), Univ. Paris-Saclay, 91191 Gif-sur-Yvette Cedex, France
2 Department of Physics "A. Pontremoli", University of Milan, 20133 Milan, Italy
3 Department of Chemical Engineering and Biotechnology, Univ. of Cambridge, Philippa Fawcett Drive, CB30AS Cambridge, U.K.; Cavendish Laboratory, University of Cambridge, JJ Thomson Avenue, CB30HE Cambridge, U.K.

It is commonly accepted that shear waves do not propagate in a liquid medium. The shear wave energy is supposed to dissipate nearly instantaneously. This statement originates from the difficulty to access "static" shear stress in macroscopic liquids. In this paper, we take a different approach. We focus on the stability of the thermal equilibrium while the liquid (glycerol) is submitted to a sudden shear strain at sub-millimetre scale. A thermal response of the deformed liquid is unveiled. The liquid exhibits simultaneous and opposite bands of about +0.04°C and -0.04°C temperature variations. The sudden thermal changes exclude the possibility of heat transfer and highlight the ability of the liquid to store the shear energy in non-uniform thermodynamic states. The thermal effects depend nearly linearly on the amplitude of the deformation supporting the hypothesis of a shear wave propagation (elastic correlations) extending up to several hundreds microns. This new physical effect can be explained in terms of the underlying phonon physics of confined liquids, which unveils a hidden solid-like response with many similarities to glassy systems.

Mechanical studies focus on the response of a sample to external forces and deformations. In stress relaxation measurements, the sample is submitted to nearly instant deformation, either in elongation or in shear geometry. The stress relaxation is used to understand the underlying mechanisms of different materials to an external mechanical field. This traditional protocol is usually applied to solids or viscoelastic solid materials, like metals, glassy polymers [1, 2], or liquid crystals [3]. For metals and polymers, a linear relationship "between inflexion slope of stress – log time curves and total decrease in stress" [1] was found. Simulations showed that the stress relaxation is similar in metals and polymers [4]. For smectic liquid crystals, relaxation times of ~ 50ms were measured and residual stress was identified [3]. For liquids with short molecular relaxation times (e.g. $\tau_{Glycerol} = 10^{-9}$s [5,6]), the shear stress is expected to dissipate nearly instantly for frequency lower than MHz or GHz excitations, making mechanically induced stress relaxation studies hypothetically irrelevant.

In this paper, we probe the response at room temperature of liquid glycerol to a constant mechanical shear strain by applying a step strain function at the sub-millimeter (100μm-250μm). In contrast with conventional stress relaxation measurements, that are believed to be irrelevant for molecular liquids, the main emphasis is given to the thermal response to mechanical deformation. The thermal radiation is a probe of the local dynamics (rotational and vibrational motions of molecules). At room temperature, the thermal radiation emits in mid-infrared wavelengths (7-14μm, ~30 THz) [7]. It is connected to the temperature using well-established thermal radiation laws (Stefan-Boltzmann law). We evidence the emergence of hot and cold non-uniform temperature bands in the liquid glycerol upon applying a step strain field. These strain induced thermodynamic regions are the non-ambiguous proof that the shear energy can be stored thermally in the liquid, leading to new solutions of its total free energy.

Glycerol (99% purity) was used for the measurements at room temperature (~22°C), far away from any critical point (e.g. for glass transition, $T_g$=-93°C). The glycerol is a much-studied liquid due its extremely wide range of uses and its biocompatibility. This glass former exhibits at room temperature a viscosity about a hundred times higher than liquid water ($\eta_{glycerol}$ = 1.41 Pa.s).

The liquid is set in the gap between α-alumina surfaces to ensure a high wetting of the liquid on the fixture (plate-plate disks of 40mm of diameter). The gap thickness varies from 100μm to 250μm. The shear strain is defined as $\gamma = \delta l / e$, where $\delta l$ is the displacement and $e$ the gap thickness. The deformation (step strain) is an almost Heaviside function $H(t)$, where $H(t) = \begin{cases} 0, & t < 0 \\ \gamma, & t > 0 \end{cases}$ (Fig.1b inset). The strain value ranges from 20% to 9500%, depending on the gap thickness (using a strain-controlled mechanical device (TA-ARESII)). The duration of the applied strain varies from 5s to 60s.

A bolometric detector of 382 x 288 pixels is used to detect thermal radiation emitted from the liquid. The experimental configuration provides information about the thermal state of the liquid along the whole gap thickness. The infra-red sensor was placed 50mm away from the edge of the plate-liquid-plate, ensuring non-contact measurements. The depth of thermal field was experimentally estimated in the glycerol at 0.85mm. The infra-red sensor captures two-dimensional thermal images with a frame rate of 27 Hz. The images were corrected by subtraction of the median value recorded during equilibrium.

Fig.1 describes the real-time thermal radiation emitted at room temperature by the glycerol at rest ($t < t_{ap}$) and upon a step (shear) strain ramp ($t \geq t_{ap}$). The colour mapping highlights the temperature variation from the thermal equilibrium (prior the measurement).

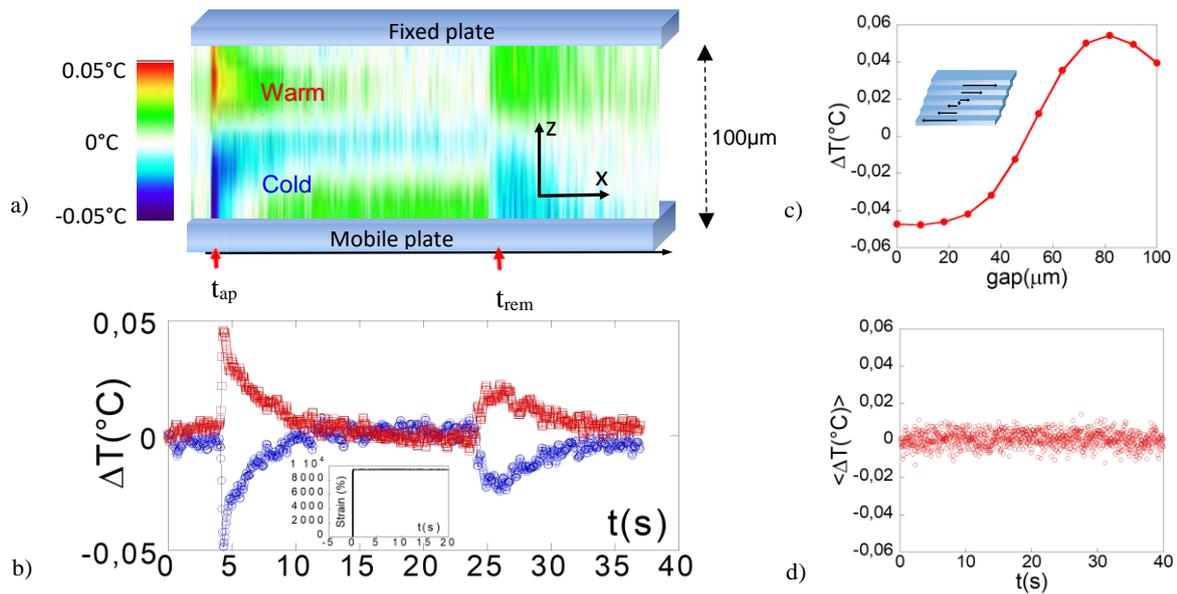

Figure 1: a) Real-time thermal mapping showing that the liquid glycerol emits a transient thermal signal by applying a step strain mechanical stimulus (shear step strain $\gamma$=9500% applied at $t_{ap}$ and removed at $t_{rem}$ ($t_{rem}$ - $t_{ap}$=20s), room temperature measurements). X-axis is the time; z-axis is along the gap thickness (100µm). The colour index indicates the temperature variation with respect to the thermal equilibrium (at $t < t_{ap}$). b) Graphical representation of Fig.1a. Top (hot) band:□, bottom (cold) band: ○. Inset: Step strain function (real data). c) Temperature variation profile along the gap (z-axis) at $t=t_{ap}+\Delta t$, averaged over three successive frames ($t_{tot}$=0.11s) just after $t_{ap}$. The insert schemes the velocity gradient viewed from the middle of the gap. d) Average temperature variation integrated over the whole gap versus time during the experiment (data of Fig.1a).

Prior the experiment (Fig.1a at $t < t_{ap}$), the liquid is in a stable equilibrium state for a long time (> 5-10 min). At $t_{ap}$, when the step shear strain (Heaviside function) is applied, an instant variation of temperature is observed. Two bands of opposite temperature are immediately created along the strain direction, disrupting the initial temperature stability of the liquid. The sudden split of the liquid in two opposite temperature variation regions, is followed by a slow thermal relaxation. The symmetry between the cold and the hot regions indicates that a thermal compensation between the two generated bands takes place simultaneously. The maximum temperature variation of the top (hot band) is + 0.04 ± 0.01°C, while for the bottom (cold) band is - 0.04 ± 0.01°C (Fig.1b). The cold band is located next to the moving plate, the source of the shear strain, while the hot band is created above the cold one. At $t_{rem}$ (Fig.1b), the applied shear strain is removed by moving the mobile surface to its initial equilibrium position ("backward" displacement). The backward motion generates a new thermal response in reaction to the displacement. This one is characterized by an amplitude half of the initial thermal response ("forward" displacement). The weaker thermal response could be attributed to non-apparent non-equilibrium conditions, where possibly created liquid-liquid interfaces reduce the transmission of stress, and hence the energy in the medium.

We now focus on the first part of the thermal liquid response ($t_{ap} < t < t_{rem}$). Fig.1c shows the profile of the temperature variation along the gap at the early instants of the step strain. The profile indicates that the hot and the cold bands are not two independent regions but follow a smooth temperature profile. Hot and cold bands are thus two sides of the same thermal effect. The middle of the gap corresponds to the temperature at rest (equilibrium temperature). Such profile might be related to the velocity gradient profile assuming the symmetry of shear strain geometry (inset of Fig.1c).

The study of the overall average temperature in the liquid gap is also instructive; the variation does not exceed $\delta T = 0.01°C$ along the whole step strain relaxation (Fig.1d). This value is within the error bar meaning that no heat exchange took place between the liquid and the environment.

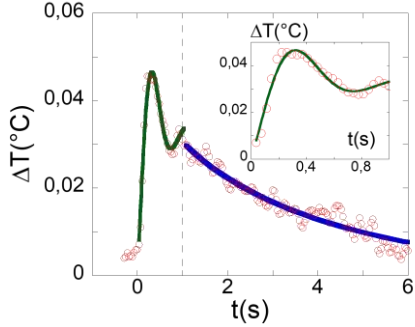

Figure 2: Modelling of the thermal response of the hot (top) band of glycerol at 0.100mm, 9500%. Green and blue lines represent the fits with Eqs. (1) and (3) respectively Inset: Detail and fit of the early times (left of the dotted line) of the thermal overshoot (green line).

Fig.2 details the evolution and the modelling of the thermal profile of the hot band (symmetrical evolution is observed for the cold one). The time to reach the maximum temperature variation is 0.19s (Fig. 2). The time of the ramp of the step strain (time needed for the strain to establish) is 0.03s; as a result, during a time estimated of 0.16s, the liquid stores the shear energy as potential (internal) energy in the liquid. This potential energy manifests as changes of macroscopic properties, such as the temperature. The duration of the ramp is the same regardless of the strain value, thus increasing the strain in each measurement is equivalent to increasing the shear rate (energy rate transferred to the liquid).

The insert of Fig.2 focuses on the early times of the thermal response of the hot (top) band, during the step strain (red circles). The temperature increases rapidly as the step strain is applied (left of dotted line), then it decreases rapidly creating an overshoot, followed by a small oscillation (second overshoot). The short time thermal response can be described by a second order transfer function $H_{nf}(s) = H_0[\omega_0^2/(s^2+2\zeta\omega_0 s+\omega_0^2)]$, where $\omega_0$ is the natural frequency, $\zeta$ is the damping ratio, $s$, a variable in the Laplace domain and $H_0$ the system gain. The mathematical solution is written as [8]:

$$\Delta T_{fit}(t) = A[1 - \frac{e^{-\zeta\omega_n t}}{\sqrt{1-\zeta^2}}\cos(\omega_n\sqrt{1-\zeta^2}t - \varphi)] \quad (1),$$

where A is a constant, φ is a phase shift and $\Delta T_{fit}$ is the temperature variation based on the second order step response. Equation (1) is fitted to the initial thermal response (Fig.2 inset). From the fitting of Eq. (1), we get: $\zeta = 0.35 \pm 0.02$, while $\omega_n = 8 \pm 0.3$ rad/s. It shows that the order of the natural frequency $\omega_n$ is of the order of Hz; i.e. describing a collective thermal effect. The characteristic overshoot is noticeable at 0.3s (Fig.2 inset), as expected for the underdamped case ($\zeta < 1$) [8]. The steady state value (asymptotic value at long timescale) is derived from the overshoot value, utilizing the equation:

$$\frac{Overshoot\ value}{Steady\ state\ value} = \exp\left(-\frac{-\zeta\pi}{\sqrt{1-\zeta^2}}\right) \quad (2)$$

[9]. The Eq. (2) foresees an asymptotic value $\Delta T_{t\to\infty} = 0.03°C$. Fig.2 shows that the thermal relaxation does not reach this value but returns to equilibrium.

The long time thermal relaxation to equilibrium (both cold and hot bands relax symmetrically) can be modelled by a stretched exponential decay $\Delta T = \Delta T_{max}\cdot(e^{-t/\tau})^\beta$ (3) (Fig.2 right of the dotted line) with the exponent estimated at $\beta \approx 0.8-0.85$ and a relaxation time $\tau$ dependent on the gap thickness and the strain value (Fig.3a). The relaxation time $\tau$ increases by decreasing the gap (for $e = 250\mu m$, $\tau$ is about ~ 0.3-0.5s, about 0.5-1.8s for $150\mu m < e < 200\mu m$, and $\tau \sim 2$-5.5s for $100\mu m$).

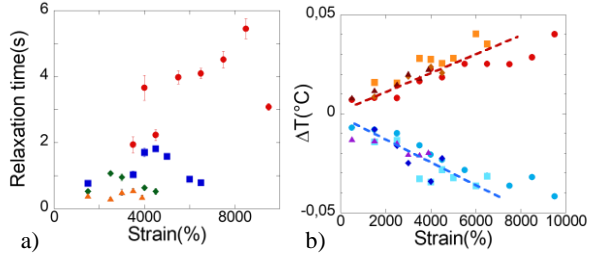

Figure 3: a) Relaxation time $\tau$(s) of the top (hot) thermal band versus shear strain amplitude. Red, blue, green and yellow points correspond to 100μm, 150μm, 200μm and 250μm gap thickness respectively. b) Amplitude of the thermal overshoot variation versus strain at different gap thicknesses. Warm band: 100μm: (●), 150μm: (■), 200μm: (♦) and 250μm: (▲). Cold band: 100μm: (●), 150μm: (■), 200μm: (♦), 250μm: (▲). Room temperature measurements.

Hence, we have seen that the temperature profiles in the system behave asymptotically like

$$\Delta T(t) \sim \exp(-\zeta \omega_n t) = \exp(-t/\tau) \quad (4)$$

where $\omega_n$ is a normal mode (eigenfrequency) of the system, and the second equality defines a characteristic relaxation time $\tau = (\zeta \omega_n)^{-1}$. According to this definition, the increase of $\tau$ upon decreasing the thickness $e$ could be explained with the decrease of damping $\zeta$ upon decreasing $e$; i.e. an increase of rigidity in agreement with the observed increase of shear elasticity at small scales [10-15].

It is well known that stretched exponential relaxation arises when there is a distribution of relaxation times in the system (e.g. from dynamical heterogeneity which is ubiquitous in glassy systems),

$$\exp(-t/\tau)^\beta \approx \int_0^\infty \rho(\tau) \exp(-t/\tau) \, d\tau \quad (5)$$

where $\rho(\tau)$ is a suitable distribution of relaxation times, which must satisfy few mathematical requirements [16]. Since $\tau$ in our system is related to the normal mode $\omega_n$ of the liquid system, the distribution $\rho(\tau)$ must be given by the distribution of vibrational normal modes in the liquid, $\rho(\omega_n)$, also known in condensed matter physics as the vibrational density of states (VDOS). Hence, upon using the VDOS of a glassy liquid in Eq. (5) we obtain [17-19]

$$\Delta T(t) \sim \exp(-t/\tau)^\beta \approx$$
$$\int_0^\infty \rho(\omega_n) \exp(-\zeta \omega_n t) \, d\omega_n \quad (6)$$

with $\tau = (\zeta \omega_n)^{-1}$. The validity of Equation (6) for the case of glycerol in the supercooled liquid state was numerically verified in Ref. [17]. It was shown that the boson peak (excess of vibrational modes over the Debye level $\sim \omega_n^2$) in the VDOS represents a crucial requirement for Eq. (6) to produce a stretched-exponential relaxation in the measured dielectric response of glycerol. Therefore, the experimentally observed stretched-exponential profile in the current system may suggest the existence of low-energy transverse phonon modes of the kind that constitute the boson peak excess of low-energy modes in glasses. In turn, this directly hints to the existence of long-range shear elastic waves in the sub-millimeter confined glycerol.

This picture based on the existence of underlying low-frequency shear modes akin to transverse phonons (the existence of which has been demonstrated also for simple liquids [20, 21]) may also explain the emergence of the cold and the hot bands following the shear strain deformation. The cold band is associated with regions where the fluid is locally expanded, whereas the hot band is associated with regions where the fluid is locally compressed. These regions are macroscopically large since they contain a huge number of molecules, hence it makes sense to define long-wavelength phonon modes that live in these regions. In the hot/compressed regions the volume V of the region gets reduced, whereas in the cold/expanded regions the volume V is enlarged. Let us recall the definition of the Grüneisen parameter: $\gamma = -\frac{d \ln \omega_n}{d \ln V}$

Since the Grüneisen parameter $\gamma$ is normally positive for liquids [22], this relation gives (upon integration) that the phonon frequency $\omega_n$ increases as the volume $V$ decreases, whereas $\omega$ decreases as the volume $V$ increases. Since the vibrational frequency is related to temperature, via $T = h\omega_n/k_B$, it is clear that $T$ increases in the regions where $V$ decreases ("compressed" regions), whereas $T$ decreases in the regions where $V$ increases ("expanded" regions), thus leading to a hot and a cold band for the compressed and the expanded regions, respectively. The same result can be obtained, by using the Anderson-Grüneisen relation [23]:

$$\gamma = -\frac{d \ln T}{d \ln V}$$

which more directly leads to the same result and applies for isentropic processes, hence also for adiabatic processes since in our case the entropy production is small since the average temperature does not change appreciably (Fig.1d).

Fig.3b shows the influence of the strain amplitude on the temperature variation in the hot and the cool parts. The evolution is symmetrical (positive and negative temperature variations) and approximately of the same magnitude for each strain value. The amplitude of the temperature variation increases nearly linearly with increasing strain (Fig.3b). The strain dependence is little influenced by the thickness for the tested gap (within 100 - 250µm) indicating that the thermal variation is mainly a bulk property. The nearly linear strain-dependence of the thermal bands defines negative and positive thermo-mechanical constants. The constants relative to the temperature $(\Delta T/T)/\gamma$ are about $\lambda_{shear} \sim -0.15\ 10^{-4}$ for the cold band and $\lambda_{shear} \sim +0.12\ 10^{-4}$ for the hot band at T=300 K. These values indicate relative variation of temperature per (shear) stretching unit. The thermo-elastic constant of steelsheets is typically about $\lambda_{elongational} = (\Delta T/T)/E \cong 0.45$ where E is the elongational strain [24]. The temperature variation is much higher in solids and is always positive in agreement with dissipative processes (moving defects). The liquid thermomechanical constants are about three decades lower indicating a much weaker energy mechanism that might be compatible with elastic intermolecular interactions [10-15].

The above results demonstrate unexpected thermo-elastic effects in liquids which can only be explained in terms of the unsuspected ability of liquids to support low-frequency shear waves. The experiments show that, upon a shear step deformation of large amplitude and at mesoscopic scale, a viscous liquid (glycerol) changes its thermodynamic state. The liquid loses its temperature homogeneity creating two major thermodynamic regions, where the temperature deviates symmetrically from the equilibrium one. These mechanically induced thermodynamic rearrangements imply an ability of the liquid to store the strain energy in its normal modes. Hot and cold thermal bands indicate that the liquid does not relax immediately: part of the shear energy is stored in the liquid, leading to the creation of local thermal bands. The temperature variations of the main two bands are of opposite values leading to an average thermal compensation and a global thermal invariance (adiabaticity). The early time thermal response (Fig.2 inset) described by a second-order response means indeed that there is an exchange of energy between two storages. One is the external shear strain and the others are viscous and elastic elements in the liquid. Equivalent systems are damped mechanical oscillators like electronic RLC circuit [8], hydro-mechanical or spring-mass-damper systems. After the overshoot, the liquid loses the gain from the step shear strain, though constant strain is maintained. This long time thermal relaxation shows a stretched exponential nature, similar to the dielectric α relaxation of glycerol near glass transition [17], elucidating a solid-like behaviour of the confined liquid glycerol. But, the timescale of the thermal relaxations is neither related to a slow relaxation heat transfer process nor to short molecular relaxation times. As the phenomenon is fast without heat transfer, the observed thermal effects (negative and positive temperature variations) are related to internal energy changes (adiabatic processes) which minimize the system free energy. This can be explained in terms of the local dilation/compression leading to local cooling/heating as a consequence of the relationship between temperature variation and volume variation established by the Grüneisen or Anderson-Grüneisen relation. In turn, the shear waves behave like phonons in solids and provide an explanation for the observed thermal bands (hot and cold in compressed and dilated regions respectively) in terms of the Grüneisen parameter relation, which is well known for phonons in solids, and relates local temperature variations to local volume variations as observed in the experiments. The formation of thermal bands is the proof that elastic shear waves propagate in liquids at the scale of hundreds microns. An elastic-like process achieved nearly without heat transfer is characteristic of adiabatic shear elasticity. The ability of liquids to support shear waves was already experimentally uncovered [10-15, 25] and theoretically suggested [20-21, 26-28]. Here the identification of a mechanically induced thermal-response allows us to demonstrate the adiabatic character of the elastic response of liquids.

These results suggest that the thermo-elastic response of liquids should be more systematically considered in order to achieve a deeper understanding of the liquid state. With confined liquid systems gaining more and more relevance, from the micro (e.g. microfluidics) to even the nano (e.g. nanofluidics) scale, this effect may play a prominent role for the manipulation of liquids at the smallest scales in future research.

We thank P. Baroni for instrumental innovation and assistance. This work has received funding from the European Union's Horizon 2020 research and innovation programme under the Marie Sklodowska-Curie grant agreement Nº 766007 and from the LabeX Palm (ANR-11-Idex-0003-02).

[1] Kuba't, J. Stress Relaxation in Solids. *Nature* **205**, 378–379 (1965).

[2] D.W.Van Krevelen, Revised by K.Te Nijenhuis. Chapter 13 - Mechanical Properties of Solid Polymers. Properties of Polymers (Fourth Edition) Their Correlation with Chemical Structure; Their Numerical Estimation and Prediction from Additive Group Contributions 2009, Pages 383-503.

[3] R. Bartolino and G. Durand. Plasticity in a Smectic-A Liquid Crystal. *Phys. Rev. Lett.* **39**, 1346 (1977).

[4] S. Blonski, W. Brostow, J. Kuba't. Molecular-dynamics simulations of stress relaxation in metals and polymers. *Phys. Rev. B* **49**, 6494 (1994).

[5] Comez, L., Fioretto, D., Scarponi, F., Monaco, G., Density fluctuations in the intermediate glass-former glycerol: a Brillouin light scattering study. *J. Chem. Phys.* **119,** 6032 (2003).

[6] Scarponi F., Comez L., Fioretto D., and Palmieri L., Brillouin light scattering from transverse and longitudinal acoustic waves in glycerol. *Phys. Rev. B* **70**, 054203 (2004).

[7] M. Vollmer, K-P Möllmann, Chapter 1 Fundamentals of Infrared Thermal Imaging. Infrared Thermal Imaging: Fundamentals, Research and Applications, Wiley-VCH Verlag GmbH & Co. KGaA (2010)

[8] M.T..Thompson. Chapter 2 - Review of Signal Processing Basics. Intuitive Analog Circuit Design. Elsevier 2014, Pages 15-52.

[9] W. Bolton, Control Systems, Pages 85-98, Elsevier (2002).

[10] B.V. Derjaguin, U.B. Bazaron, K.T. Zandanova, O.R. Budaev. The complex shear modulus of polymeric and small-molecule liquids. *Polymer* **30**(1), 97 (1989). https://doi.org/10.1016/0032-3861(89)90389-3.

[11] L. Noirez, H. Mendil-Jakani, P. Baroni, Identification of finite shear-elasticity in the liquid state of molecular and polymeric glass-formers. *Phil. Mag.* **91** 1977–1986 (2011). http://dx.doi.org/10.1080/14786435.2010.536176.

[12] L. Noirez, P. Baroni. Revealing the solid-like nature of glycerol at ambient temperature. *J. of Mol. Struct.* **972**, 16-21 (2010).

[13] L. Noirez, P. Baroni. Identification of a low-frequency elastic behaviour in liquid water. *J Phys Condens Matter* **24** (37), 372101 (2012). https://doi.org/10.1088/0953-8984/24/37/372101

[14] P. Lv, Z. Yang, Z. Hua, M. Li, M. Lin, Z. Dong. Viscosity of water and hydrocarbon changes with micro-crevice Thickness. *Colloids and Surfaces A: Physicochemical and Engineering Aspects* **504**, 287 (2016). https://doi.org/10.1016/j.colsurfa.2016.05.083.

[15] Y. Chushkin, C. Caronna, A. Madsen, Low-frequency elastic behavior of a supercooled liquid. *Europhys. Letters* **83** 36001 (2008).

[16] D. C. Johnston, Stretched exponential relaxation arising from a continuous sum of exponential decays, Phys. Rev. B 74, 184430 (2006).

[17] B. Cui, R. Milkus, A. Zaccone, Direct link between boson-peak modes and dielectric α-relaxation in glasses, Phys. Rev. E 95, 022603 (2017).

[18] B. Cui, R. Milkus, A. Zaccone, The relation between stretched-exponential relaxation and the vibrational density of states in glassy disordered systems, *Phys. Lett. A* **381**, 446–451 (2017).

[19] A. Zaccone, Relaxation and vibrational properties in metal alloys and other disordered systems, *J. Phys.: Condens. Matter* **32,** 203001 (2020).

[20] C. Yang, M. T. Dove, V. V. Brazhkin, & K. Trachenko. Emergence and Evolution of the k Gap in Spectra of Liquid and Supercritical States, *Phys. Rev. Lett.* **118**, 215502 (2017).

[21] A. Zaccone, K Trachenko. Explaining the low-frequency shear elasticity of confined liquids. arXiv:2007.11916.

[22] L. Knopoff and J. N. Shapiro, Pseudo-Grüneisen Parameter for Liquids, Phys. Rev. B 1, 3893 (1970).

[23] O. L. Anderson, The Grüneisen Parameter for the Last 30 Years, Geophys. J. Int. 143, 279 (2000).

[24] R. Munier, C. Doudard, S. Calloch, B. Weber, Determination of high cycle fatigue properties of a wide range of steelsheet grades from self-heating measurement, Inter. J. of Fatigue, **63** 46 (2014).

[25] L. Noirez, P. Baroni, J. F. Bardeau, Highlighting non-uniform temperatures close to liquid/solid surface. *Appl. Phys. Lett.* **110**, 213904 (2017). https://doi.org/10.1063/1.4983489.

[26] F. Volino, Théorie visco-élastique non-extensive. *Ann. Phys.* Fr 22(1-2):7–41 (1997).

[27] K. Trachenko, Lagrangian formulation and symmetrical description of liquid dynamics. *Phys. Rev. E* **96**, 062134 (2017) doi: https://doi.org/10.1103/PhysRevE.96.062134

[28] M. Baggioli, V. Brazhkin, K. Trachenko, M. Vasin. Gapped momentum states, *Physics Reports* **865**, 1-44 (2020).